\documentclass[twocolumn,english,nofootinbib]{revtex4}
\usepackage[T1]{fontenc}
\usepackage[latin9]{inputenc}
\usepackage{amsmath}
\usepackage{amssymb}
\usepackage{graphicx}

\makeatletter

\providecommand{\tabularnewline}{\\}

\@ifundefined{textcolor}{}
{%
 \definecolor{BLACK}{gray}{0}
 \definecolor{WHITE}{gray}{1}
 \definecolor{RED}{rgb}{1,0,0}
 \definecolor{GREEN}{rgb}{0,1,0}
 \definecolor{BLUE}{rgb}{0,0,1}
 \definecolor{CYAN}{cmyk}{1,0,0,0}
 \definecolor{MAGENTA}{cmyk}{0,1,0,0}
 \definecolor{YELLOW}{cmyk}{0,0,1,0}
}

\usepackage{babel}

\usepackage{ulem}

\usepackage{babel}

\usepackage{babel}

\makeatother

\usepackage{babel}
\begin{document}

\title{Gravitational wave emission from oscillating millisecond pulsars}

\author{Mark G. Alford and Kai Schwenzer}

\address{Department of Physics, Washington University, St. Louis, Missouri,
63130, USA}
\begin{abstract}
Neutron stars undergoing r-mode oscillation emit gravitational radiation
that might be detected on earth. For known millisecond pulsars the
observed spindown rate imposes an upper limit on the possible gravitational
wave signal of these sources. Taking into account the physics of r-mode
evolution, we show that only sources spinning at frequencies above
a few hundred Hertz can be unstable to r-modes, and we derive a more
stringent \textit{universal r-mode spindown limit} on their gravitational
wave signal, exploiting the fact that the r-mode saturation amplitude
is insensitive to the structural properties of individual sources.
We find that this refined bound limits the gravitational wave strain
from millisecond pulsars to values below the detection sensitivity
of next-generation detectors. Young sources are therefore a more promising
option for the detection of gravitational waves emitted by r-modes
and to probe the interior composition of compact stars in the near
future. 
\end{abstract}
\maketitle

\section{Introduction}

With the significant sensitivity improvement of forthcoming next generation
gravitational wave detectors like advanced LIGO \cite{Harry:2010zz},
advanced Virgo \cite{Weinstein:2011kh} and the LCGT \cite{Kuroda:2010zzb}
there is a realistic chance that gravitational waves may be directly
observed. In addition to transient events such as neutron star and/or
black hole mergers or supernovae, which require that an event happens
sufficiently near to us during the observation period \cite{Abadie:2012rq},
it is important to also consider continuous sources. Millisecond pulsars
are a particularly promising class since they are very old and stable
systems and therefore could be reliable sources of gravitational waves.
Their fast rotation strongly favors gravitational wave emission \cite{Aasi:2013sia},
and the fact that their timing behavior is known to high precision
\cite{Manchester:2004bp} greatly simplifies the analysis required
to find a signal in the detector data. Emission due to deformation
of these objects (``mountains''), which is usually parametrized
by an ellipticity, is the standard paradigm for continuous gravitational
wave searches \cite{Collaboration:2009rfa}, but global oscillation
modes of a star can also emit copious gravitational waves that could
be detectable if the oscillation reaches sufficiently high amplitudes.
R-modes are the most interesting class \cite{Andersson:1997xt,Andersson:2000mf,Lindblom:1998wf,Owen:1998xg},
because they are generically unstable in millisecond pulsars and therefore
will be present unless the dissipative damping is strong enough.

If r-modes arise in a spinning neutron star, they affect the spindown
(since they cause the star to lose angular momentum via gravitational
radiation) and the cooling (since the the damping forces on the r-mode
generate heat). To understand the interplay of these effects we have
developed \cite{Alford:2012yn,Alford:2013pma} an effective description
of the spindown evolution where complicated details about the star's
interior are absorbed into a few effective parameters. The resulting
spindown can be rather different from that predicted by simpler approaches,
and includes strict bounds on the uncertainties in the final results.
In this paper we will use this method to analyze the possible r-mode
gravitational radiation of old neutron stars. Firstly, however, we
provide some background and motivation.

R-modes can occur in young or old pulsars. In the case of young sources
\cite{Aasi:2013sia,Abadie:2010hv,Abbott:2008fx,Wette:2008hg} we have
analyzed their r-mode evolution \cite{Alford:2012yn} and found that
r-modes can provide a \textit{quantitative} explanation for their
observed low spin rates. Moreover, the r-mode gravitational emission
is expected to be strong, because a large r-mode amplitude would be
required to spin down the known young pulsars to their current low
spin frequencies within their lifetimes which are as short as a thousand
years. These known pulsars are no longer in their r-mode spindown
epoch, but there may be unobserved young neutron stars, e.g. associated
with known supernova remnants such as SN 1987A, that are currently
undergoing r-mode spindown, and several of them would be in the sensitivity
range of advanced LIGO \cite{Alford:2012yn}, allowing this scenario
to be falsified by future measurements.

In this paper we focus on old neutron stars which have been spun up
by accretion, and we perform an analysis of their expected r-mode
gravitational wave radiation. In \cite{Alford:2013pma} novel r-mode
instability regions in spindown timing parameter space have been derived
that allow us to decide if r-modes can be present in old millisecond
radio pulsars. As discussed there, there are two scenarios to explain
the observed timing data. It might be that the ordinary nuclear matter
model of neutron stars is incomplete, and there is additional damping
(e.g.~from exotic forms of matter or currently overlooked physical
processes) that stops r-modes from growing in these stars. In this
case there will be no r-mode gravitational radiation from old neutron
stars. The other possibility is the conventional scenario where only
standard damping mechanisms are present in neutron stars. In this
scenario most old millisecond pulsars \textit{will be} undergoing
r-mode oscillations, since for expected r-mode saturation amplitudes
the dissipative heating ensures that fast spinning sources can neither
cool nor spin out of the parameter region were r-modes are unstable
\cite{Alford:2013pma}. Yet, some slower spinning sources can escape
the instability region and we will determine the limiting frequency.
Therefore, there will be gravitational radiation from most old neutron
stars in this scenario, and the purpose of this paper is to find out
whether it could be detected on earth.

The detectability of known continuous sources is generally described
by the ``spindown limit'' which is, for a specific source with known
timing data, the maximum gravitational wave strain that can be emitted
by that source. Despite the quite restrictive limits set by the spindown
data, the large spin frequencies of millisecond pulsars could nevertheless
lead to a detectable signal. Present gravitational wave detectors---like
the original LIGO interferometer---did not probe the spindown limit
for millisecond pulsars. However, next generation detectors including
the advanced LIGO detector will be able to beat the spindown limit
for various sources. Therefore, it is interesting to assess the chance
to detect gravitational emission from oscillating millisecond pulsars.

We will introduce here the \textit{universal r-mode spindown limit}
on the gravitational wave strain, which is more restrictive since
it takes into account our understanding of the r-mode spindown and
the complete information we have about these systems. Whereas deformations
of a given source depend on its evolutionary history and could therefore
vary significantly from one source to another, for proposed saturation
mechanisms the r-mode saturation amplitude proves to be rather insensitive
to details of a particular source, like its mass or radius \cite{Bondarescu:2013xwa,Alford:2011pi,Haskell:2013hja}.
The expected gravitational wave strain of a given source can then
be strongly constrained by the timing data of the \textit{entire set}
of millisecond pulsars. Using our semi-analytic approach to pulsar
evolution, and assuming that the same saturation and cooling mechanism
(with given power-law dependence on temperature) operates in all the
stars, we can then obtain the universal limit given in Eq.~(\ref{eq:universal-spindown-limit}).
We will see that this is considerably below the standard spindown
limits, indicating that it will be harder than previously expected
to see r-mode gravitational waves from these sources.

\section{R-mode spindown of millisecond pulsars\label{sec:R-mode-spindown-of}}

As described in \cite{Alford:2012yn,Alford:2013pma} the r-mode evolution
\cite{Owen:1998xg} can be discussed within an effective description,
which relies on the fact that a compact star appears effectively as
a point source and that the relevant material properties integrated
over the star have simple power law dependencies on the macroscopic
observables that change during the evolution. The relevant macroscopic
quantities are the power emitted as gravitational waves $P_{G}$,
the dissipated power $P_{D}$ that heats the star and the thermal
luminosity $L$ that cools the star 
\begin{equation}
P_{G}=\hat{G}\Omega^{8}\alpha^{2},\; P_{D}=\hat{D}T^{\delta}\Omega^{\psi}\alpha^{\phi},\; L=\hat{L}T^{\theta}\,,\label{eq:powers}
\end{equation}
in terms of the rotational angular velocity $\Omega=2\pi f$, the
core temperature $T$ of the star and the dimensionless r-mode amplitude
$\alpha$ defined in \cite{Lindblom:1998wf,Alford:2012yn}. The explicit
form of the prefactors $\hat{G}$, $\hat{D}$ and $\hat{L}$ for different
damping and cooling mechanisms \cite{Alford:2013pma} is given in
tab. \ref{tab:parameterization}.

\begin{table}
\begin{tabular}{|c||c|}
\hline 
parameter of the ...  & integral expression\tabularnewline
\hline 
\hline 
GW luminosity  & $\hat{G}\equiv\frac{2^{17}\pi}{3^{8}5^{2}}\tilde{J}^{2}GM^{2}R^{6}$\tabularnewline
\hline 
\hline 
Shear visc. dissipation  & $\hat{D}=5\tilde{S}\Lambda_{{\rm QCD}}^{3+\sigma}R^{3}$\tabularnewline
\hline 
Bulk visc. dissipation  & $\hat{D}=\frac{2^{3}}{3^{3}7}\frac{\Lambda_{{\rm QCD}}^{9-\delta}\tilde{V}R^{7}}{\Lambda_{{\rm EW}}^{4}}$\tabularnewline
\hline 
Ekman layer dissipation  & $\hat{D}=5\left(\frac{2}{3}\right)^{\frac{9}{2}}\frac{3401+2176\sqrt{2}}{11!!}\sqrt{\hat{\eta}_{c}\rho_{c}}R_{c}^{4}$\tabularnewline
\hline 
\hline 
Neutrino luminosity  & $\hat{L}=\frac{4\pi R^{3}\Lambda_{{\rm QCD}}^{9-\theta}\tilde{L}}{\Lambda_{{\rm EW}}^{4}}$\tabularnewline
\hline 
Photon luminosity  & $\hat{L}=\frac{\pi^{3}}{15}R^{2}\hat{X}^{4}$\tabularnewline
\hline 
\end{tabular}\caption{\label{tab:parameterization}Parameters in the general parameterization
eq.~(\ref{eq:powers}) for the energy loss rates. The arising quantities
are the mass $M$ and the radius $R$ of the star, the gravitational
constant $G$, generic normalization scales $\Lambda_{{\rm QCD}}$
and $\Lambda_{{\rm EW}}$ and in case of Ekman damping the relevant
quantities at the crust/core interface, see \cite{Lindblom:2000gu}.
The dimensionless constants $\tilde{J}$, $\tilde{V},$ $\tilde{S}$
and $\tilde{L}$ contain the complete information about the interior
of the star. Their definition and values for realistic neutron stars
are given in \cite{Alford:2012yn}.}
\end{table}

\begin{figure}
\includegraphics{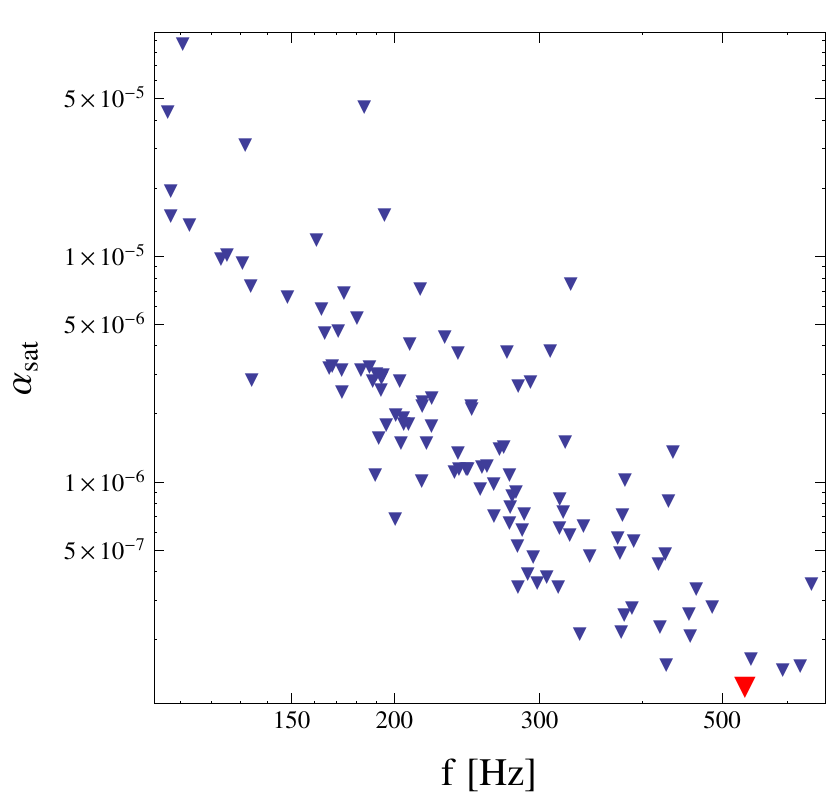}

\caption{\label{fig:alpha-bounds} Upper bounds on the r-mode saturation amplitude
arising from the observed spindown of the pulsars in the ATNF database
\cite{Manchester:2004bp}. The strongest bound $\alpha_{{\rm sat}}\lesssim1.2\times10^{-7}$
is obtained for the $533$ Hz pulsar J0034-0534 which has a spindown
rate $\dot{f}\approx-1.4\cdot10^{-15}s^{-2}$ (large triangle at lower
right, red online). }
\end{figure}

R-modes are unstable and their fast growth has to be stopped by a
non-linear dissipative saturation mechanism. Even though there are
several interesting proposals \cite{Bondarescu:2013xwa,Haskell:2013hja,Alford:2011pi,Lin:2004wx,Lindblom:2000az,Wu:2000qy,Rezzolla:1999he}
it is not yet settled which mechanism will dominate and saturate r-modes.
For millisecond pulsars we expect moderate saturation amplitudes,
in which case the pulsar spindown is determined by the equation \cite{Owen:1998xg}

\begin{equation}
\frac{d\Omega}{dt}=-\frac{3\hat{G}\alpha_{{\rm sat}}^{2}\left(T,\Omega\right)}{I}\Omega^{7}-\cdots\,,\label{eq:spindown}
\end{equation}
in terms of the moment of inertia of the star $I$ and the r-mode
saturation amplitude $\alpha_{{\rm sat}}$. The observed total spindown
rate will in general be larger since in addition to r-modes there
are other spindown mechanisms given by the ellipsis. Nevertheless,
by assuming that the observed spindown rate is entirely due to r-modes,
observed pulsar timing data allows one to give upper bounds on the
r-mode saturation amplitude. These bounds are shown for the observed
radio pulsars included in the ATNF database \cite{Manchester:2004bp}
in fig.~\ref{fig:alpha-bounds} and they require very low saturation
amplitudes, $10^{-7}\lesssim\alpha_{{\rm sat}}\lesssim10^{-5}$ \cite{Alford:2013pma,Bondarescu:2013xwa}.
Similar low values are obtained from pulsars in low mass x-ray binaries
\cite{Alford:2013pma,Mahmoodifar:2013quw}. Moreover, one can see
in fig.~\ref{fig:alpha-bounds} that faster spinning sources generally
set more stringent bounds on the saturation amplitude of r-modes in
the considered pulsar.

The r-mode saturation amplitude can in general depend both on the
temperature and the frequency of the star. We use a general parametrization
of the saturation amplitude with a power-law form 
\begin{equation}
\alpha_{{\rm sat}}\left(T,\Omega\right)=\hat{\alpha}_{{\rm sat}}T^{\beta}\Omega^{\gamma}\,,\label{eq:saturation-amplitude}
\end{equation}
as realized for the proposed saturation mechanisms \cite{Bondarescu:2013xwa,Haskell:2013hja,Alford:2011pi,Wu:2000qy}.
Here the exponents are fixed (rational) numbers determined by the
saturation mechanism, whereas the reduced amplitude $\hat{\alpha}_{{\rm sat}}$
is less well known and can also depend on parameters of the particular
source, like the mass or the radius. Using this general approach it
was found in \cite{Alford:2012yn} that the r-mode heating is significant
even for small amplitude modes and the thermal evolution is systematically
faster than the spindown. Therefore the star reaches a thermal steady
state where the dissipative r-mode heating balances the cooling due
to photons and neutrinos and the temperature is given by

\begin{equation}
T_{{\rm hc}}=\left(\frac{\hat{G}\hat{\alpha}_{{\rm sat}}^{2}\Omega^{8+2\gamma}}{\hat{L}}\right)^{{\textstyle \frac{1}{\theta-2\beta}}}\,.\label{eq:heating-cooling}
\end{equation}
This leads to a spindown equation along this steady state curve \cite{Alford:2012yn,Alford:2013pma}

\begin{equation}
\frac{d\Omega}{dt}=-\frac{3\hat{G}^{\theta/\left(\theta-2\beta\right)}\hat{\alpha}_{{\rm sat}}^{2\theta/\left(\theta-2\beta\right)}}{I\hat{L}^{2\beta/\left(\theta-2\beta\right)}}\Omega^{n_{rm}}\,,\label{eq:effective-spindown}
\end{equation}
with an effective braking index

\begin{equation}
n_{{\rm rm}}=7\left(\!\frac{1\!+\!2\gamma/7\!+\!2\beta/\left(7\theta\right)}{1\!-\!2\beta/\theta}\!\right)\label{eq:effective-braking-index}
\end{equation}
that depends on the saturation mechanism and can be rather different
from the generic r-mode spindown exponent $7$.

The spindown equation has the solution

\begin{align}
\Omega\!\left(t\right) & =\left((\Omega_{i})^{z}+\frac{3z}{2}\left(\frac{\hat{G}^{\theta}\hat{\alpha}_{{\rm sat}}^{2\theta}}{\hat{L}^{2\beta}}\right)^{{\textstyle \frac{1}{\theta-2\beta}}}\frac{t-t_{i}}{I}\!\right)^{1/z}\:,\label{eq:spindown-solution}\\
z & \equiv\frac{2(3+\gamma)\theta+4\beta}{\theta-2\beta}\:.\nonumber 
\end{align}
This solution has two limits. At early times the first term in the
parenthesis of eq.~(\ref{eq:spindown-solution}) dominates, so that
the star hardly spins down, i.e. $\Omega\approx\Omega_{i}$, and at
late times the second term dominates, so that the spindown becomes
independent of the initial angular velocity $\Omega_{i}$. The crossover
point between these two regimes is determined by the reduced saturation
amplitude $\hat{\alpha}_{{\rm sat}}$. For young pulsars the $\Omega_{i}$-independent
late time behavior of the spindown law is relevant. For some old millisecond
pulsars the spindown rate is so low that the ``early time'' regime
is realized and they hardly spin down even over their billion year
age.

\begin{figure}
\includegraphics{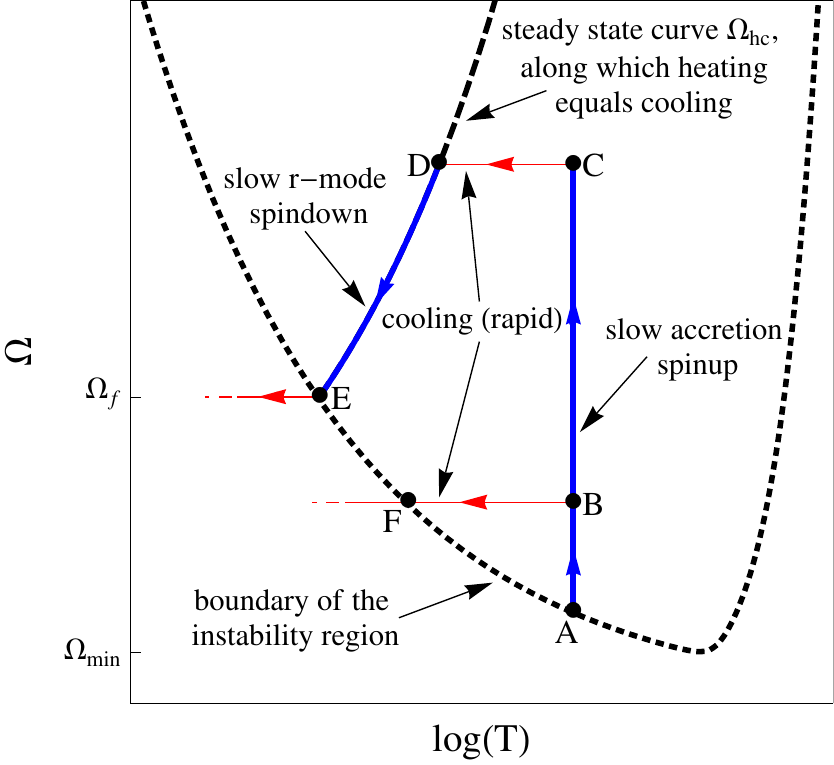}

\caption{\label{fig:schematic evolution} Schematic evolution of recycled radio
pulsars which have been spun up by accretion. Whereas the cooling
(horizontal segments) takes less than a million years, the slow spindown
along the steady state curve takes longer than a billion years. The
occurrence of a long r-mode spindown epoch is determined by whether
the frequency when accretion ends is below (B) or above (C) the universal
r-mode frequency bound $\Omega_{f}$. See the text for details.}
\end{figure}

The r-mode evolution eq.~(\ref{eq:spindown-solution}) takes place
unless the dissipation, which depends strongly on temperature and
frequency, is strong enough to completely damp r-modes. R-modes are
only unstable at sufficiently high frequencies: a typical instability
region for a neutron star with standard damping mechanisms in a $T-\Omega$-diagram
\cite{Lindblom:1998wf}{} is shown in fig.~\ref{fig:schematic evolution}.
By ``standard damping'' we mean established mechanisms%
\footnote{A potential Ekman layer at the crust-core boundary \cite{Lindblom:2000gu}
does not qualitatively change this picture \cite{Alford:2013pma}.%
}, namely shear viscosity due to leptonic and hadronic scattering \cite{Shternin:2008es}
and bulk viscosity due to modified Urca reactions \cite{Sawyer:1989dp}.
Ref.~\cite{Alford:2010fd} gives a general semi-analytic expression
for the minimum frequency $\Omega_{{\rm min}}$ down to which r-modes
can be unstable, and shows that this limit is extremely insensitive
to unknown details of the source and the microphysics (see also \cite{Lindblom:1998wf}).
Fig.~\ref{fig:schematic evolution} also shows two qualitatively
different evolution trajectories. A recycled millisecond pulsar entering
the instability region at point A in fig.~\ref{fig:schematic evolution}
is slowly spun up and kept warm by accretion in a binary system, following
the thick vertical line. The r-mode evolution, eq.~(\ref{eq:spindown-solution}),
starts when accretion stops. This may occur when the star is spinning
slowly (B) or quickly (C). In either case, even though the star is
in the region of $\Omega$-$T$ space where according to the standard
damping mechanism r-modes are unstable, the star then cools faster
than it spins down (following the thin horizontal lines). If accretion
brought the star to a high spin frequency (C) then the star cools
until it reaches the steady state line (dashed line, given by eq.~(\ref{eq:heating-cooling}))
at point (D); it then slowly spins down, following the steady-state
line, and would only reach the boundary of the instability (E) after
time scales that are longer than the age of known sources. Therefore
we expect such a source not too far below point (D). Fig.~\ref{fig:schematic evolution}
shows a steady-state line for low r-mode saturation amplitude, in
which case the line is high enough that it exits the instability region
at a frequency $\Omega_{f}$ which is significantly above the minimum
frequency $\Omega_{{\rm min}}$. This means that if accretion leaves
the star with a low spin frequency (B), below $\Omega_{f}$, then
the star cools in less than a million years \cite{Yakovlev:2004iq}
and reaches the boundary of the instability region (F). The value
of $\Omega_{f}$ is 
\begin{equation}
\Omega_{f}=\left(\frac{\hat{D}^{\theta-2\beta}\hat{\alpha}_{{\rm sat}}^{2\delta}}{\hat{G}^{\theta-\delta-2\beta}\hat{L}^{\delta}}\right)^{\frac{1}{6\theta-8\delta-12\beta-2\delta\gamma}}\ .\label{eq:final-frequency}
\end{equation}
It was shown in \cite{Alford:2012yn,Alford:2013pma} that this expression
is extremely insensitive to the microphysical details. Whereas for
young sources discussed in \cite{Alford:2012yn} neutrino emission
is the relevant cooling mechanism to determine the final spindown
frequency, for the low saturation amplitudes relevant for millisecond
pulsars photon cooling from the surface and damping due to shear viscosity
dominates \cite{Alford:2013pma} in eq.~(\ref{eq:final-frequency}).
For a given upper bound on $\hat{\alpha}_{{\rm sat}}$, all sources
below this \textit{universal r-mode frequency bound} $\Omega_{f}$
cannot be undergoing r-mode spindown since they either have been spun
out of the instability region or cooled out of it in less than a million
years, which is considerably shorter than their billion year age.
In contrast, all sources spinning faster than $\Omega_{f}$ must be
undergoing r-mode spindown (i.e.~they are on the steady-state curve
in fig.~\ref{fig:schematic evolution}). The fastest spinning sources
($f\gtrsim600\,{\rm Hz}$) could have only left the instability region
if the saturation amplitude would be as low as $\alpha_{{\rm sat}}\lesssim10^{-10}$
\cite{Alford:2013pma}, which is orders of magnitude below what proposed
saturation mechanisms can provide \cite{Bondarescu:2013xwa,Haskell:2013hja,Alford:2011pi,Wu:2000qy}.
We conclude that fast spinning sources should be emitting gravitational
waves via r-mode spindown and we will determine the required spin
frequencies below.

\section{Gravitational wave strain}

\begin{figure}
\includegraphics{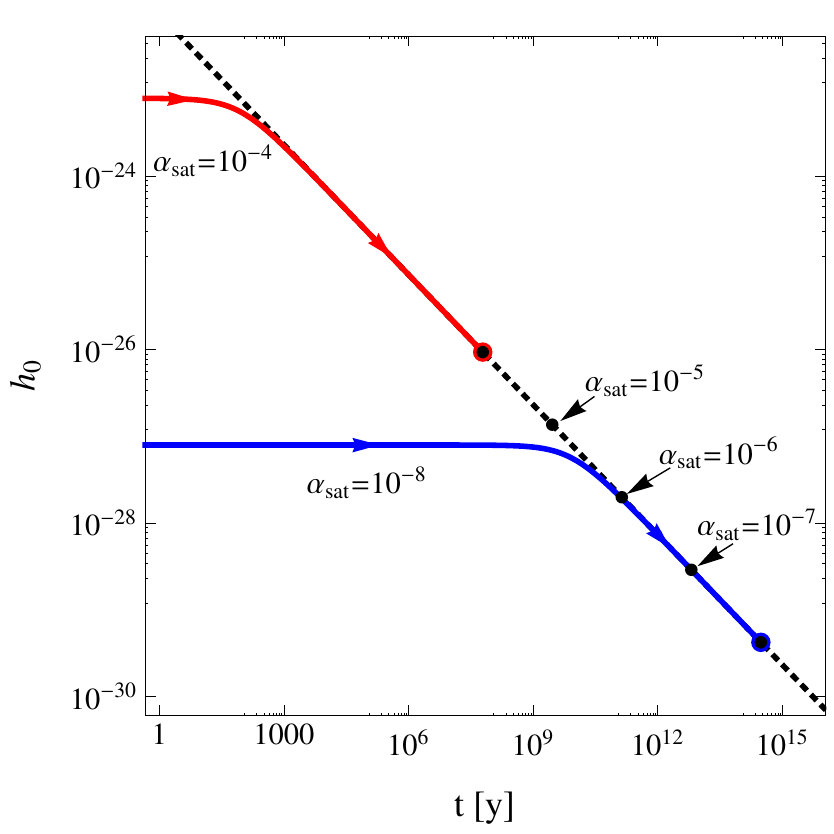}

\caption{\label{fig:strain-evo}The time evolution of the emitted intrinsic
gravitational wave strain amplitude (solid curves) and the endpoints
of the gravitational wave emission (dots) \cite{Alford:2012yn} shown
for different saturation amplitudes and for a fiducial source located
at a distance of $1\,{\rm kpc}$. The universal late time behavior
is also shown (dotted line).}
\end{figure}

R-modes emit gravitational waves due to their time-varying current
quadrupole moment. The gravitational wave frequency $\nu$ emitted
by the dominant fundamental ($m=2)$ r-mode is related to the rotational
angular velocity via $\nu=2/\left(3\pi\right)\Omega$ \cite{Owen:1998xg}.
The gravitational wave signal of a given source is described by the
intrinsic gravitational wave strain of the detector, which describes
the expected signal in a terrestrial detector and can directly be
compared to the detector noise. For r-modes it takes the form \cite{Owen:2010ng}

\begin{equation}
h_{0}=\sqrt{\frac{2^{15}\pi^{7}}{5}}\frac{\tilde{J}GMR^{3}\nu^{3}\alpha_{{\rm sat}}}{D}\:,\label{eq:intrinsic-strain-amplitude}
\end{equation}
where $D$ is the distance to the source. In a recent study of the
gravitational wave emission of young sources \cite{Alford:2012yn}
it was found that for the large amplitudes required to explain the
low spin frequencies of young pulsars, the late time behavior of the
spindown evolution eq.~(\ref{eq:spindown-solution}) is relevant
and in this case the strain eq.~(\ref{eq:intrinsic-strain-amplitude})
depends only on the age and the distance of the source \cite{Wette:2008hg},
and is independent of the saturation amplitude. Because of the restrictive
bounds on the saturation amplitude shown in fig.~\ref{fig:alpha-bounds},
for some old radio pulsars the ``early time'' limit of the evolution
is relevant, where the frequency barely changes and the strain depends
linearly on the saturation amplitude. In fig.~\ref{fig:schematic evolution}
such a source stays close to its starting point D on the spindown
curve for more than a billion years. However, in general the time
evolution is relevant and has to be taken into account. This can be
seen in fig.~\ref{fig:strain-evo} where the evolution of the gravitational
wave strain is shown for saturation amplitudes relevant for millisecond
pulsars. The dots also show for various amplitudes the end of the
gravitational wave emission, where the source spins slowly enough
that the r-mode is damped. As seen, for $\alpha_{{\rm sat}}<10^{-5}$
the time needed for a star to spin out of the r-mode instability region
is considerably more than a billion years, longer than the age of
these sources.

\subsection{Standard r-mode spindown limit}

\begin{figure}
\includegraphics{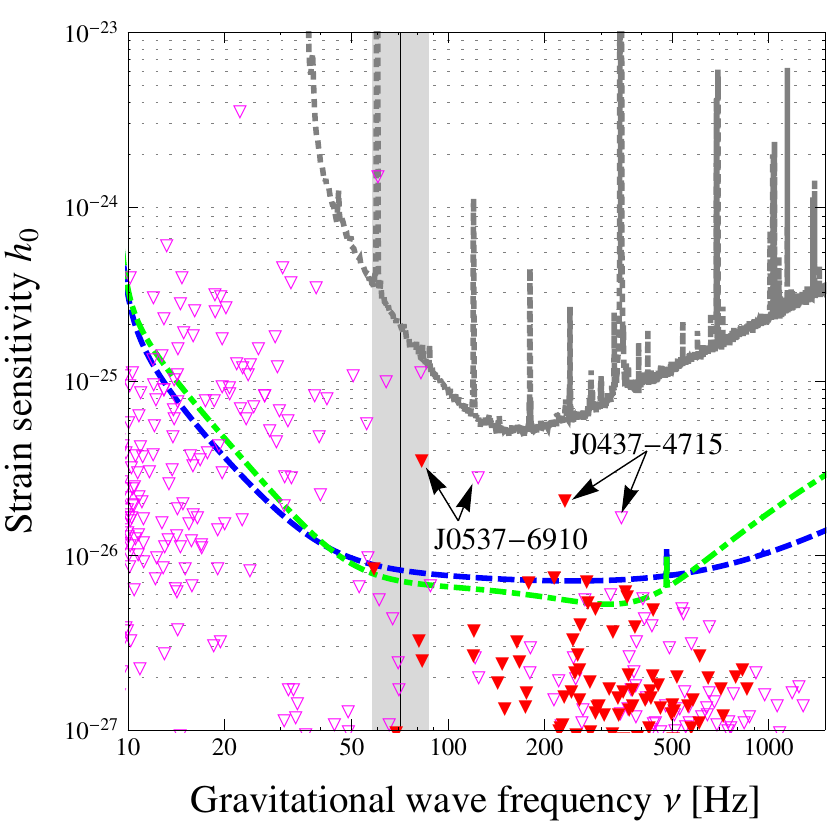}

\caption{\label{fig:spindown-limits}The standard spindown limits of known
radio pulsars compared to the characteristic strain amplitude for
different detector configurations assuming a coherent analysis of
a year of data. Open (magenta) triangles show the limits for the standard
case of elliptic deformations of the star \cite{Aasi:2013sia} and
filled (red) triangles for the case of r-mode gravitational wave emission.
The solid (grey) curve gives the sensitivity of the original LIGO
detector and the dashed (blue) and dot-dashed (green) curves show
the sensitivity of the advanced LIGO detector in the standard and
neutron star enhanced mode. The vertical line shows the limiting frequency
below which r-modes are absent in a neutron star and the shaded band
gives the uncertainty on it using the semi-analytic result \cite{Alford:2010fd}
and the ranges for the underlying parameters used in \cite{Alford:2012yn}.}
\end{figure}

Using the spindown equation (\ref{eq:spindown}) to eliminate the
r-mode saturation amplitude $\alpha_{{\rm sat}}$ in eq.~(\ref{eq:intrinsic-strain-amplitude}),
i.e. employing the values given in fig.~\ref{fig:alpha-bounds},
yields the \textit{spindown limit}%
\footnote{The expression given here is slightly smaller than the estimate given
in \cite{Owen:2010ng}, since the rotational energy loss goes not
entirely into gravitational waves but is partly dissipated to saturate
the r-mode at a finite amplitude.%
} \cite{Owen:2010ng,Aasi:2013sia}

\begin{equation}
h_{0}^{\left({\rm sl}\right)}=\sqrt{\frac{15}{4}\frac{GI\left|\dot{\Omega}\right|}{D^{2}\Omega}}\:,\label{eq:spindown-limit-strain}
\end{equation}
which provides an upper bound that is saturated when the entire rotational
energy loss is due to the gravitational wave emission and accompanying
dissipation caused by r-modes. The spindown limits for the observed
radio pulsar data \cite{Manchester:2004bp} are shown in fig.~\ref{fig:spindown-limits}
and are represented by inverted triangles. Here the spindown limits
from r-modes (solid triangles) are compared to those for the typically
considered case of elliptic star deformations (open triangles), which
has recently been studied in detail by the LIGO collaboration \cite{Aasi:2013sia}.
R-modes lead to a slightly higher strain but at a lower frequency
\cite{Owen:2010ng}. These limits are also compared to the detector
sensitivity at 95\% confidence limit (curves), given by $h_{0}^{95\%}\!\approx\!10.8\sqrt{S_{h}/\Delta t}$
\cite{Aasi:2013sia} in terms of the spectral density $S_{h}$ of
the detector strain noise and the observation interval $\Delta t$.
To assess fig.~\ref{fig:spindown-limits} it is important to recall
that r-modes are only unstable at sufficiently large frequencies.
The lowest frequency at which r-modes are unstable ($\Omega_{{\rm min}}$
in Fig.~\ref{fig:schematic evolution}) \cite{Alford:2010fd,Lindblom:1998wf}
shown by the vertical line, sets a strict frequency limit below which
no r-mode gravitational wave emission is possible.

Despite this restriction, the figure shows that the spindown limit
for several millisecond radio pulsars should be beaten by the advanced
LIGO detector. However, even though the pulsars J0537-6910 and J0437-4715
could be significantly above the detector sensitivity they are not
promising sources. The pulsar J0537-6910 is actually a young pulsar
that has been analyzed in detail in \cite{Alford:2012yn} where it
is shown that although it is slightly above the minimum frequency
of the instability region, it is very likely outside of it. The $f=174\,{\rm Hz}$
pulsar J0437-4715 is the closest and brightest millisecond pulsar
and therefore would be a natural target. However, this is also the
only non-accreting source for which a temperature estimate is available
\cite{Haskell:2012}, and this shows that it is outside of the instability
region and similarly cannot emit gravitational waves due to r-modes.
Moreover, its frequency is below likely values of $\Omega_{f}$ (eq.~(\ref{eq:final-frequency}))
so, as we discussed at the end of section \ref{sec:R-mode-spindown-of}
and will analyze in more detail below, it ought to have already cooled
out of the instability region for a neutron star with standard damping
mechanisms \cite{Alford:2013pma}. Thinking beyond the advanced LIGO
sensitivity thresholds shown in fig.~\ref{fig:spindown-limits},
we note that planned detectors like the Einstein telescope, which
has an order of magnitude higher sensitivity, would be able to detect
the gravitational waves that would be emitted from many sources, if
r-modes are responsible for the better part of their observed spindown
rate. 

\subsection{Universal r-mode spindown limit}

The spindown limit for a particular source only takes into account
information about that source. Here we will derive a more restrictive
limit taking into account the entire data set of radio pulsars. It
is based on the observation that proposed r-mode saturation mechanisms
are very insensitive to the details of a particular source \cite{Bondarescu:2013xwa,Alford:2011pi,Haskell:2013hja}.
To make this statement quantitative, we factorize the reduced saturation
amplitude given in eq.~(\ref{eq:saturation-amplitude}) by writing
\begin{equation}
\hat{\alpha}_{{\rm sat}}=\hat{\alpha}_{{\rm sat}}^{\left({\rm mic}\right)}\hat{\alpha}_{{\rm sat}}^{\left({\rm mac}\right)}\:,
\end{equation}
where $\hat{\alpha}_{{\rm sat}}^{\left({\rm mic}\right)}$ depends
on the microphysics of the saturation mechanism and is source-independent,
and $\alpha_{{\rm sat}}^{\left({\rm mac}\right)}$ depends on the
macroscopic properties of a specific source (mass, radius, etc) %
\footnote{To make the factorization unambiguous, we will use the convention
that once a set of macroscopic parameters have been chosen, the source-dependent
macroscopic part $\alpha_{{\rm sat}}^{\left({\rm mac}\right)}$ consists
of powers of those parameters with no multiplicative prefactor. For
the saturation mechanisms considered here there is a known set of
macroscopic parameters (mass, radius, etc) but for generality we do
not limit ourselves by writing $\alpha_{{\rm sat}}^{\left({\rm mac}\right)}$
explicitly in terms of them.%
}, which generically only vary within narrow margins. For a given saturation
mechanism determined by $\beta$ and $\gamma$ in eq.~(\ref{eq:saturation-amplitude})
and a particular source-dependence encoded in $\hat{\alpha}_{{\rm sat}}^{\left({\rm mac}\right)}$
we can then use eq.~(\ref{eq:effective-spindown}) to determine the
reduced microscopic saturation amplitudes $\hat{\alpha}_{{\rm sat}}^{\left({\rm mic}\right)}$
from given pulsar timing data. The smallest value obtained from the
entire data set, which is realized for a particular source with frequency
$f_{0}$ and spindown rate $\dot{f}_{0}$, can then be used to give
a limit for a general source, spinning with frequency $f$, using
eqs.~(\ref{eq:heating-cooling}), (\ref{eq:effective-spindown})
and (\ref{eq:intrinsic-strain-amplitude}). We find the \textit{universal
r-mode spindown limit}

\begin{equation}
h_{0}^{\left({\rm usl}\right)}=\sqrt{\frac{15}{4}\frac{GI\left|\dot{f}_{0}\right|}{D^{2}f_{0}}}\left(\frac{\hat{\alpha}_{{\rm sat}}^{\left({\rm mac}\right)}}{\hat{\alpha}_{{\rm sat},0}^{\left({\rm mac}\right)}}\right)^{{\textstyle \frac{1}{1-2\beta/\theta}}}\left(\frac{f}{f_{0}}\right)^{{\textstyle \frac{3+\gamma+2\beta/\theta}{1-2\beta/\theta}}}\:.\label{eq:universal-spindown-limit}
\end{equation}
The first factor is just the standard spindown limit eq.~(\ref{eq:spindown-limit-strain})
for the source with the strongest bound on the reduced microscopic
saturation amplitude, whereas the two others are correction factors
involving information on the source to which this limit applies, with
exponents $\beta$, $\gamma$, $\theta$ from eqs.~(\ref{eq:powers})
and (\ref{eq:saturation-amplitude}). Note that this result is independent
of the details of the cooling mechanism encoded in $\hat{L}$ although
the effective spindown law eq.~(\ref{eq:effective-spindown}) that
was used to obtain eq.~(\ref{eq:universal-spindown-limit}) depends
on it. It depends on the power law exponent $\theta$ of the cooling
luminosity via the factor $2\beta/\theta$, which takes different
values depending on whether the cooling is dominated by neutrinos
($\theta=8$ for modified Urca cooling) or photons ($\theta=4$).
Recently it has been shown \cite{Alford:2013pma} that unless the
dissipation is so strong that r-modes are completely damped away,
radio pulsars should be surprisingly hot due to the strong heating
from r-mode dissipation. Therefore, both photon and neutrino cooling
can be relevant for observed radio pulsars. Since the detailed properties
of particular sources are generally unknown the ratio of the macroscopic
parts of the reduced saturation amplitudes in eq.~(\ref{eq:universal-spindown-limit})
can only be estimated. However, from our theoretical understanding
of compact stars (possible range of masses, radii, etc) we can, for
a given saturation mechanism, determine bounds on this unknown factor
that are tight enough that the universal spindown limit is still considerably
more restrictive than the standard spindown limit.

\begin{figure}
\includegraphics{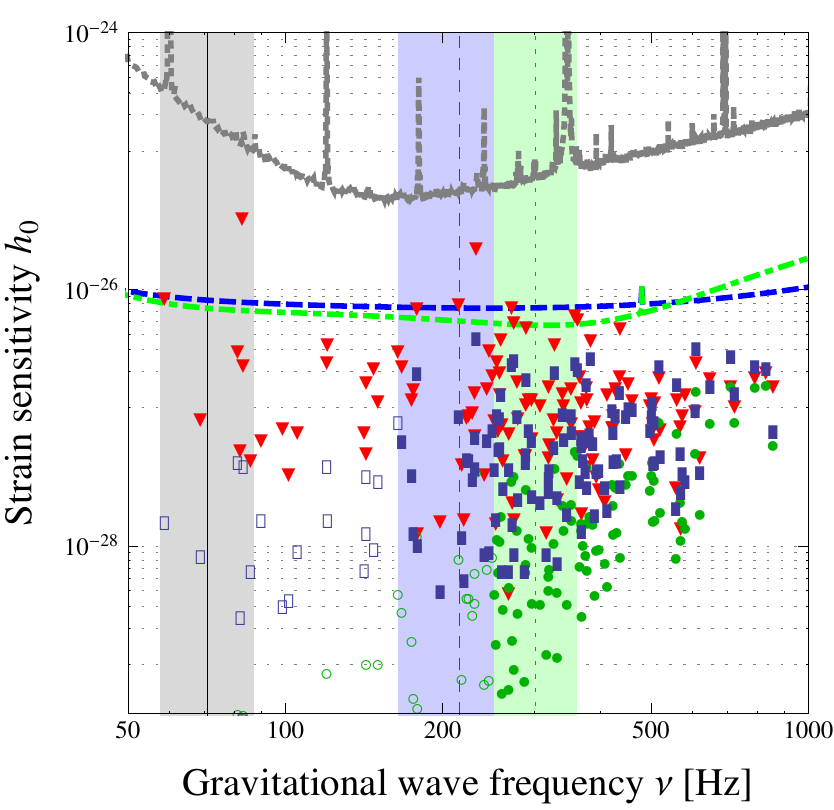}

\caption{\label{fig:universal-spindown-limits}Comparison of different upper
bounds on the strain amplitude of known radio pulsars due to r-mode
emission. The spindown limit (red triangles) is obtained from the
timing data of an individual source. The universal spindown limit
takes into account that the saturation mechanism applies to the entire
class of millisecond pulsars, and also provides a lower bound eq.~(\ref{eq:final-frequency})
on the frequency. We show universal spindown limits (green circles)
and minimum frequency (green dotted vertical line with uncertainty
band) for the toy model of a constant r-mode saturation amplitude
\cite{Owen:1998xg}. We also show universal spindown limits (blue
rectangles) and minimum frequency (blue dashed vertical line with
uncertainty band) for a realistic saturation mechanism arising from
mode-coupling and the damping of the daughter modes by shear viscosity
\cite{Bondarescu:2013xwa}. For a given saturation mechanism, stars
below the minimum frequency (open symbols) do not undergo r-mode oscillation.}
\end{figure}

The simplest and most often used toy model for r-mode saturation
\cite{Owen:1998xg} assumes a constant saturation amplitude that is
independent of both temperature and frequency, so $\beta=\gamma=0$.
Although realistic models based on an explicit physical saturation
mechanism have a more complicated dependence this simple case is useful
for illustrative purposes. In this case the saturation amplitude is
also assumed to be independent of the source so $\alpha_{{\rm sat}}=\hat{\alpha}_{{\rm sat}}^{\left({\rm mic}\right)}$,
which is given in fig.~\ref{fig:alpha-bounds}. The strongest limit
$\alpha_{{\rm sat}}\leq1.2\times10^{-7}$ is obtained for the fast
pulsar J0034-0534 with $f_{0}\approx533\,{\rm Hz}$ and $\dot{f}_{0}\approx-1.4\times10^{-15}\,{\rm s}^{-1}$.
Using this bound in eq.~(\ref{eq:final-frequency}) shows that in
the constant saturation model r-mode gravitational wave emission can
only be present in sources spinning with frequencies $f\gtrsim225\,{\rm Hz}$
corresponding to gravitational wave frequencies $\nu\gtrsim300\,{\rm Hz}$,
since slower spinning sources would have left the r-mode instability
region (see fig.~\ref{fig:schematic evolution} and the accompanying
discussion). The expression for the universal spindown limit shows
that the bounds for other sources scale in this case as $\left(f/f_{0}\right)^{3}$.
Therefore the universal spindown limits are significantly lower than
the standard spindown limits since the saturation amplitude obtained
from the entire data set is lower and the frequencies of most sources
are lower than $f_{0}$. This is shown in fig.~\ref{fig:universal-spindown-limits}
which compares the universal spindown limits for the constant saturation
amplitude model (circles) to the standard spindown limits (triangles)
given before in fig.~\ref{fig:spindown-limits}. The, dashed vertical
line in fig.~\ref{fig:universal-spindown-limits} gives the universal
r-mode frequency bound eq.~\ref{eq:final-frequency} below which
r-modes cannot be present. Therefore slower spinning sources, which
appeared to be rather promising when only taking into account the
standard spindown limit, are entirely excluded, as is denoted by the
open symbols. But even for faster spinning sources the universal spindown
limits can be orders of magnitude smaller. Therefore, all limits for
this saturation mechanism are considerably below the estimated sensitivity
of advanced LIGO.

R-mode saturation amplitudes obtained from realistic mechanisms have
a temperature and/or frequency dependence. Moreover, the power law
exponents that are found are generally negative and of order one.
As an important realistic example we discuss the saturation due to
mode coupling and the subsequent damping of the daughter modes \cite{Arras:2002dw}.
The saturation amplitude from mode coupling has recently been revised
\cite{Bondarescu:2013xwa}, taking into account that the dominant
damping source for daughter modes in a neutron star is likely shear
viscosity instead of the previously assumed boundary layer damping.
The revised saturation amplitude could be low enough to be compatible
with the restrictive bounds from the observed small spindown rates
given in fig.~\ref{fig:alpha-bounds}. In the case of a star with
an impermeable crust the saturation amplitude is given by \cite{Bondarescu:2013xwa}

\begin{equation}
\alpha_{{\rm sat}}=\frac{\left|C_{R}\right|}{\sqrt{\tilde{J}}}=\left(1.4\times10^{-7}\frac{K_{4}^{\frac{2}{3}}}{\kappa_{D}}\right)\left(\frac{1}{\sqrt{\tilde{J}}R_{10}^{\frac{4}{3}}}\right)T_{8}^{-\frac{4}{3}}f_{500}^{-\frac{2}{3}}\:.\label{eq:mode-coupling-alpha}
\end{equation}
Here the first parenthesis represents $\hat{\alpha}_{{\rm sat}}^{\left({\rm mic}\right)}$,
which can be determined from the spindown data. The lowest bound on
$\hat{\alpha}_{{\rm sat}}^{\left({\rm mic}\right)}$ is obtained from
the $f_{0}=336\,{\rm Hz}$ radio pulsar J2229+2643. Using this bound
in eq.~(\ref{eq:final-frequency}), in the mode coupling model r-modes
are only present in sources that spin with frequencies $f\gtrsim160\,{\rm Hz}$
corresponding to gravitational wave frequencies $\nu\gtrsim215\,{\rm Hz}$.
In the mode-coupling mechanism the saturation amplitude has a strong
temperature and density dependence which gives a weaker scaling of
the universal spindown limit. As seen in eq.~(\ref{eq:universal-spindown-limit})
the scaling ranges from $\left(f/f_{0}\right)^{3/2}$ if neutrino
cooling dominates to $f/f_{0}$ if photon cooling is dominant. To
obtain rigorous upper bounds we are conservative and use for each
source the weaker of the two constraints obtained from neutrino and
photon cooling. The second parenthesis in eq.~(\ref{eq:mode-coupling-alpha})
is the source-dependent factor $\hat{\alpha}_{{\rm sat}}^{\left({\rm mac}\right)}$
which is unknown. To estimate the uncertainty in this factor, we note
that the radius of a neutron star $R_{10}$ (in units of $10\,{\rm km}$)
is at present uncertain within $1\lesssim R_{10}\lesssim1.5$ and
the factor $\tilde{J}$, defined in \cite{Owen:1998xg}, has been
shown in \cite{Alford:2012yn} to be strictly bounded within $1/\left(20\pi\right)\leq\tilde{J}\leq3/\left(28\pi\right)$.
Therefore, the uncertainty on the universal spindown limit from the
source-dependent factor in eq.~(\ref{eq:universal-spindown-limit})
is $\hat{\alpha}_{{\rm sat}}^{\left({\rm mac}\right)}/\hat{\alpha}_{{\rm sat,0}}^{\left({\rm mac}\right)}\lesssim2.5$
Including this uncertainty, the results for the universal spindown
limit for saturation due to mode-coupling are shown as well in fig.~\ref{fig:universal-spindown-limits}
(rectangles) and the corresponding frequency below which r-modes are
excluded is shown by the dashed vertical line. As can be seen the
universal spindown limits are above those for the constant saturation
model (circles), but in most cases still significantly below both
the standard spindown limits and for all sources they are below the
sensitivity of the advanced LIGO detector. For some fast spinning
sources the standard spindown limit is more restrictive but these
are far below the aLIGO sensitivity.

\section{Conclusions}

We have analyzed the continuous gravitational wave emission of millisecond
radio pulsars due to r-modes. As an improvement to the usual bound,
given by the spindown limit, we have derived the \textit{universal
r-mode spindown limit} which takes into account the fact that proposed
r-mode saturation mechanisms are insensitive to the macroscopic star
configuration (mass, radius, moment of inertia, etc) and takes into
account the whole class of sources. Using this additional information,
we find that the universal spindown limit for the intrinsic gravitational
wave strain amplitude can be significantly smaller than the usual
spindown limit. Furthermore, we show that r-modes are damped in old
millisecond radio pulsars spinning with frequencies below about $150-200\,{\rm Hz}$
so that corresponding gravitational wave emission is not expected
to be present. Our results do not rely on explicit estimates for the
r-mode saturation amplitude, which depend on the microphysics and
are still very uncertain, but merely on the parametric temperature
and frequency dependence which is generic for a given saturation mechanism
and given by characteristic rational power-law exponents. We compare
our improved bounds to the detection thresholds for realistic searches
with next generation detectors like advanced LIGO using a year of
coherent data and find that for none of the known millisecond radio
pulsars would r-mode gravitational waves be detectable in the near
future. This is in contrast to r-mode emission from young sources,
where several potential sources are in reach of advanced LIGO \cite{Alford:2012yn}.
However, if the sensitivity could be improved by the combination of
different detectors, the analysis of larger coherent data sets or
other enhancements, the universal spindown limits of selected millisecond
pulsars which are close to the the detection limit might be beaten
by next generation searches. For third generation detectors, like
the planned Einstein telescope, there is a realistic chance to detect
dozens of sources and our refined bounds identify those that are most
promising. In contrast to the conventional spindown limit (triangles
in fig.~\ref{fig:universal-spindown-limits}), which is sizable for
some lower frequency sources, the universal spindown limit (circles
or rectangles in fig.~\ref{fig:universal-spindown-limits}) shows
that it is actually the mid- to high-frequency sources which feature
the largest bounds.

The universal spindown limit relies on the assumption that the same
r-mode saturation mechanism is operating in the entire set of radio
pulsars. In principle it cannot be completely excluded that different
saturation mechanisms are at work in different sources. This could
happen if there were classes of sources with qualitatively different
structural or phase compositions. For the recycled old radio pulsars
that we focus on in this paper, this is not a likely scenario. These
stars are very stable systems that hardly change over time and have
very similar properties. For instance the magnetic fields that could
distinguish different radio pulsars are all rather small and are not
expected to strongly affect the r-mode evolution besides the additional
magnetic spindown. However, it is quite possible that old and young
stars have different saturation mechanisms, in fact such a difference
is required to explain the low spin frequencies of young pulsars \cite{Alford:2012yn}
and the low spindown limits of old radio pulsars \cite{Alford:2013pma}.
One possibility is enhanced dissipation due to the transformation
of a neutron star or its core into a quark star owing to the density
increase during the initial spindown \cite{Alford:2013rea}. Another
option would be enhanced dissipation in a superfluid/superconductor
\cite{Haskell:2013hja} which is only present below the superfluid
melting temperature and this transition might have been explicitly
observed in the cooling of the neutron star in Cassiopeia A \cite{Page:2010aw}.
Both of these transitions would happen in the dynamic early evolution
of young sources, before they are a few hundreds of years old. In
contrast, recycled old radio pulsars are very stable systems that
hardly change over time and have very similar properties. For instance
the magnetic fields that could distinguish different radio pulsars
are all rather small and are not expected to strongly affect the r-mode
evolution besides the additional magnetic spindown. Therefore, the
assumption that the same saturation mechanism is realized in the entire
class of old millisecond radio pulsars is reasonable.

In addition to the exciting prospect of directly detectable gravitational
waves, the emission from oscillating pulsars presents a unique chance
to directly probe the interior of a compact star. The amplitude of
r-modes, which is encoded in the gravitational wave signal, can directly
reveal the damping properties of the matter inside the star and thereby
its composition. In addition to thermal measurements from low mass
x-ray binaries \cite{Lindblom:1998wf,Haskell:2012} and pulsar timing
data \cite{Alford:2013pma,Manchester:2004bp}, gravitational waves
would provide a third messenger to probe the interior star composition
via r-modes. The combined analysis of these different data sets would
provide a clearer picture of the star's interior and could allow us
to discriminate different star compositions in the future. 
\begin{acknowledgments}
This research was supported in part by the Offices of Nuclear Physics
and High Energy Physics of the U.S. Department of Energy under contracts
\#DE-FG02-91ER40628 and \#DE-FG02-05ER41375. 
\end{acknowledgments}

\end{document}